\documentclass[aps,prl,twocolumn,groupedaddress]{revtex4}
\usepackage{graphicx}

\begin{document}


\title{Two Transitions in the Damping of a Unitary Fermi Gas}


\author{J. Kinast, A. Turlapov, and J. E. Thomas}
\email[jet@phy.duke.edu]{}
\affiliation{Duke University, Department of Physics, Durham, North
Carolina, 27708, USA}


\date{\today}

\begin{abstract}
We measure the temperature dependence of the radial breathing mode
in an optically trapped, strongly-interacting Fermi gas of $^6$Li,
just above the center of a broad Feshbach resonance. The frequency
remains close to the unitary hydrodynamic value, while the damping
rate reveals transitions at two well-separated temperatures,
consistent with the existence of atom pairs above a superfluid
transition.
\end{abstract}

\pacs{313.43}

\maketitle

Optically-trapped, unitary Fermi
gases~\cite{OHaraScience,AmScientist} test predictions for exotic
systems, from nuclear matter~\cite{Heiselberg,Baker,Carlson} and
quark-gluon plasmas~\cite{Heinz} to high temperature
superconductors~\cite{Levin}. Recent measurements of the  heat
capacity of a unitary gas reveal a
transition~\cite{KinastScience,JointScience}. This has been
interpreted as the onset of superfluidity,  using a
Bardeen-Cooper-Schrieffer~(BCS) -- Bose-Einstein Condensate~(BEC)
crossover approach which was initially developed for
high-temperature superconductors~\cite{ChenScience,JointScience}.
In a unitary Fermi gas, pair interactions between particles are
``strong'' in the sense that the zero-energy scattering length is
much greater than the interparticle spacing, as achieved by tuning
near a Feshbach resonance~\cite{OHaraScience}. Such a gas exhibits
universal
features~\cite{OHaraScience,Heiselberg,HoUniversalThermo}. At
sufficiently low temperatures, unitary Fermi gases are believed to
comprise normal atoms, condensed pairs and noncondensed
pairs~\cite{ChenScience,JointScience}. Fermionic atom pairs are
probed in  recent projection
experiments~\cite{Jincondpairs,Ketterlecondpairs,KetterleMagnetSweep}
and in measurements of the pairing gap~\cite{GrimmGap,Jingap}.
Evidence for superfluid hydrodynamics in a unitary gas appears in
anisotropic expansion~\cite{OHaraScience} and in the breathing
mode frequencies and damping
rates~\cite{Kinast,Bartenstein,KinastMagDep}.

In this Letter, we report the precision measurement of the
temperature dependence of the frequency and damping rate for the
radial breathing mode of a unitary Fermi gas of $^6$Li. We
identify two transitions in the damping rate which occur
 at well-separated temperatures. This is consistent with a qualitative picture  of a unitary gas in which
the superfluid transition temperature
 lies well below the temperature at which noncondensed pairs first form~\cite{JointScience,ChenScience}.
Remarkably, neither transition is accompanied by an abrupt change
of the frequency, which in the whole range of temperatures remains
within a few percent of the unitary hydrodynamic value. Below the
lower transition temperature, the damping rate extrapolates to
zero at zero temperature, as expected for a  superfluid. The
higher temperature transition can be interpreted as arising from
the breaking of noncondensed atom pairs by the collective
excitation.


In the experiments, we prepare a degenerate 50-50 mixture of the
two lowest spin states of $^6$Li atoms by forced
evaporation~\cite{OHaraScience} in an ultrastable CO$_2$ laser
trap~\cite{OHaraStable}.  At a bias magnetic field $B$ of 840 G,
just above the center of the Feshbach
resonance~\cite{BartensteinFeshbach,SchunckFeshbach}, the trap
depth is lowered by a factor of $\simeq 580$ in a few
seconds~\cite{OHaraScience,Kinast} and then  recompressed to 4.6\%
(for most of the experiments) of the full trap depth in 1.0 s and
held for 0.5 s to assure equilibrium. A controlled amount of
energy is added to the gas by releasing the atoms from the trap
for a short  time and then recapturing the
cloud~\cite{KinastScience,JointScience}. The gas is then allowed
to thermalize for 0.1 s.

The radial breathing mode is excited by releasing the cloud and
recapturing the atoms after $25\,\mu$s (for 4.6\% trap depth).
After the excitation, we let the cloud oscillate for a variable
time $t_{hold}$, at the end of which the gas is released and
imaged after $\simeq 1$ ms of
expansion~\cite{OHaraScience,JointScience,Kinast}.

Radial breathing mode frequencies $\omega$ and damping times
$\tau$ are determined from the oscillatory dependence of the
released cloud size on
$t_{hold}$~\cite{Kinast,Bartenstein,KinastMagDep}. For each
temperature, 60-90 values of $t_{hold}$ are chosen in the time
range of interest. These values of $t_{hold}$ are randomly ordered
during data acquisition to avoid systematic error. Three full
sequences are obtained and averaged. The averaged data is fit with
a damped sinusoid $x_{0}+A\exp(-t/\tau)\sin(\omega t +\varphi )$.
We have obtained oscillation curves at 30 different temperatures,
containing data from 6300 repetitions of the experiment.

For most of the data reported, the total number of atoms is
$N=2.0(0.2)\times 10^5$. From the measured trap frequencies,
corrected for anharmonicity, we obtain for 4.6\% trap depth:
$\omega_\perp=\sqrt{\omega_x\omega_y} = 2\pi\times 1696(10)$ Hz,
$\omega_x/\omega_y=1.107(0.004)$, and $\omega_z=2\pi\times 71(3)$
Hz, so that
$\bar{\omega}=(\omega_x\omega_y\omega_z)^{1/3}=2\pi\times 589(5)$
Hz is the mean oscillation frequency and $\lambda
=\omega_z/\omega_\perp=0.045$ is the anisotropy parameter. The
typical Fermi temperature $T_F=(3 N)^{1/3}h\bar{\omega}/k_B$ of a
corresponding noninteracting gas is $\simeq 2.4\,\mu$K, small
compared to the final trap depth of $U_0/k_B=35\,\mu$K (at 4.6\%
of full depth). The coupling parameter of the strongly-interacting
gas at $B=840$ G is  $k_Fa\simeq -30.0$, where $\hbar
k_F=\sqrt{2m\, k_B T_F}$ is the Fermi momentum, and $a=a(B)$ is
the zero-energy scattering length estimated from
Ref.~\cite{BartensteinFeshbach}.

The dimensionless empirical temperature $\tilde{T}$ is determined
by the method implemented in~\cite{KinastScience,JointScience}:
The column density of the cloud is spatially integrated in the
axial direction to yield a normalized (integrates to 1), one
dimensional, transverse spatial distribution $n(x)$. This
distribution is fit to determine the empirical reduced temperature
$\tilde{T}$ using a finite temperature Thomas-Fermi profile with a
fixed Fermi radius, which is measured in a separate experiment at
the lowest temperatures~\cite{KinastScience,JointScience}. The
empirical temperature $\tilde{T}$ is numerically calibrated to the
theoretical reduced temperature
$T/T_F$~\cite{ChenScience,JointScience}. In
Ref.~\cite{JointScience}, we show that a simple approximation
relating $\tilde{T}$ to $T/T_F$ is given by,
\begin{equation}
\tilde{T}\simeq \tilde{T}_{nat}\equiv\frac{T}{T_F\sqrt{1+\beta}}.
\label{eq:NaturalTemp}
\end{equation}
Eq.~\ref{eq:NaturalTemp} yields accurate values of  $T/T_F$ for
$\tilde{T} \geq 0.45$ and provides a reasonable estimate at lower
temperatures, where higher precision can be obtained using the
calibration~\cite{JointScience}. Here $\beta$ is the unitary gas
parameter~\cite{OHaraScience,Heiselberg,MechStab,Carlson,Strinati},
which we recently measured to be $\beta=-0.49(0.04)$ (statistical
error only)~\cite{KinastScience,JointScience}.

\begin{figure}[htb]
\includegraphics[width=3.5in]{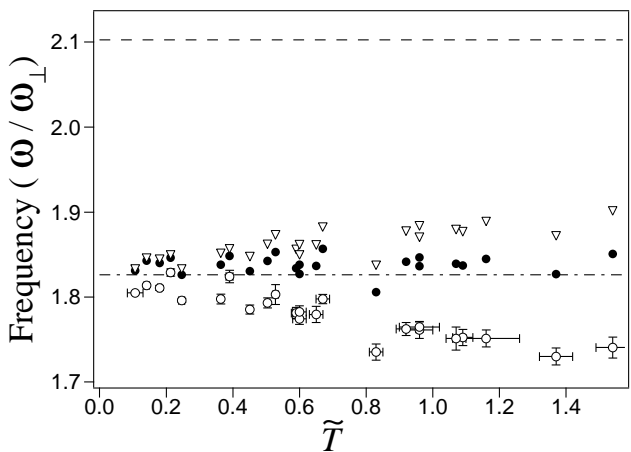}
\caption{Frequency $\omega$ versus empirical reduced temperature
$\tilde{T}$: Raw data (empty circles with error bars); Data
corrected for anharmonicity: Black dots-- using a zero-temperature
Thomas-Fermi profile  and Triangles-- using a finite-temperature
Thomas-Fermi profile. The dot-dashed line is the unitary
hydrodynamic frequency $\omega_H=\sqrt{10/3}\,\omega_\perp$. The
dashed line at the top of the scale is the frequency $2\omega_x$
observed for a noninteracting gas at the lowest temperatures.}
\label{fig:Frequency}
\end{figure}

The mode frequency provides important information on the state of
the system. The frequency versus the empirical temperature for
experiments at 4.6\% trap depth is shown in
Fig.~\ref{fig:Frequency}.
The figure shows the measured frequencies $\omega_{\mbox{meas}}$
(open circles), uncorrected for anharmonicity in the trapping
potential, as well as the frequencies after correction by two
different methods. The frequency correction is proportional to the
ratio $\langle \rho^4\rangle/\langle
\rho^2\rangle$~\cite{Stringarishift}, where $\rho$ is the
transverse radius of the expanded cloud. For an isentropic unitary
gas, we obtain~\cite{z2x2}
\begin{equation}
\omega=\omega_{\mbox{meas}}\,\left(1+\frac{2}{5}\,\frac{m\,\omega_\perp^2}{U_0}\,
\frac{\langle\rho^4\rangle}{\langle \rho^2\rangle\,b_H^2}\right),
\label{eq:FreqCorrection}
\end{equation}
with $b_H$  a hydrodynamic scale
factor~\cite{Menotti,OHaraScience}. For 1 ms of expansion, we
obtain $b_H=13.3$.

 The first method for estimating $\langle
\rho^4\rangle/\langle \rho^2\rangle$ assumes that the spatial
distribution of the gas is nearly a zero-temperature Thomas-Fermi
profile, which is a good approximation at the lowest temperatures.
In this case, the corrected frequency
is~\cite{Stringarishift,z2x2}
\begin{equation}
\omega=\omega_{\mbox{meas}}\,\left(1+\frac{32}{25}\,\frac{m\,\omega_\perp^2}{U_0}\,
\frac{\langle x^2\rangle}{b_H^2}\right), \label{eq:CorrectionTF}
\end{equation}
where $\langle x^2\rangle$ is the transverse mean square width of
the gas in the x-direction after the expansion. Applying this
method over the whole temperature range, we find that the
corrected frequencies (black dots in Fig.~\ref{fig:Frequency})
remain very close to the unitary, hydrodynamic value, shown as a
dot-dashed line.

In the second method, we include the effects of the  finite
temperature on the spatial profile of the cloud by calculating the
ratio $\langle \rho^4\rangle/\langle \rho^2\rangle =(4/3)\,\langle
x^4\rangle/\langle x^2\rangle$ directly from the fitted
one-dimensional finite temperature Thomas-Fermi profiles. The
corrected frequencies are displayed in Fig.~\ref{fig:Frequency} as
open triangles. The higher order corrections have been calculated
and are found to be negligible.

The radial breathing frequency varies smoothly over the whole
temperature range, and remains close to the value
$\omega_H=\sqrt{10/3}\,\omega_\perp=1.83\,\omega_\perp$ predicted
by hydrodynamic theory for a unitary gas, where
$1/(k_Fa)=0$~\cite{Stringariosc,Heiselbergosc,Hu,Zubarevosc1,Zubarevosc2,Manini}.
Such temperature independence has been observed previously in a
BEC~\cite{DalibardBreathing}. For the unitary Fermi gas, the
observed frequencies are far from $2\omega_x=2.10\,\omega_\perp$,
the value observed for a noninteracting gas at the lowest
temperatures, which is shown as a dashed line at the top of
Fig.~\ref{fig:Frequency}. This, as well as the observed
hydrodynamic expansion, justifies the use of the hydrodynamic
expansion factor $b_H$ in Eqs.~\ref{eq:FreqCorrection}
and~\ref{eq:CorrectionTF} over the whole temperature range.

The frequencies obtained using the finite temperature corrections
(triangles in Fig.~\ref{fig:Frequency}) rise 4\% above $\omega_H$
at the highest temperatures. This can be explained by the slow
decrease in the collision rate of the unitary gas at higher
temperatures~\cite{UnitaryCollisions}, which makes the gas
slightly less hydrodynamic and pulls the frequency up toward the
noninteracting gas value. The slow increase in frequency is
consistent with our previous estimate of the reduced temperature
for ballistic expansion of the unitary gas, $T/T_F\geq
3$~\cite{OHaraScience,UnitaryCollisions}.

\begin{figure}[htb]
\includegraphics[width=3.5in]{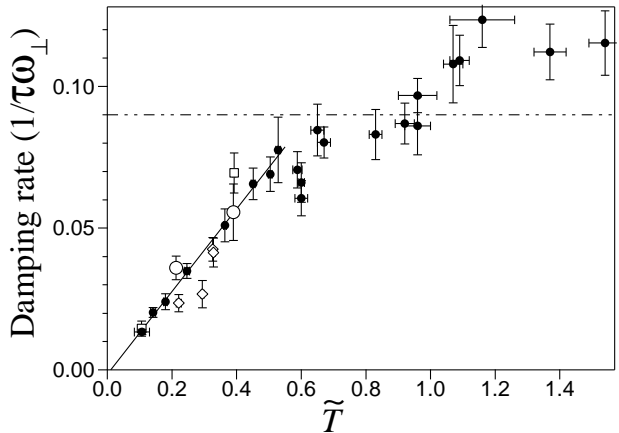}
\caption{Temperature dependence of the damping rate for the radial
breathing mode of a trapped $^6$Li gas at 840 G, showing two
transitions. Solid dots are the main data set taken at 4.6\% trap
depth. Other symbols are for the system with scaled parameters:
Two squares -- at 0.85\% of full trap depth; Four diamonds -- at
19\% of full trap depth; Two open circles~--~at 3 times smaller
number of atoms. The dot-dashed line is the maximum damping rate
for a classical harmonically trapped gas with binary collisions.
The solid line is Eq.~(\ref{eq:ExpLinearDamping}) which
extrapolates close to zero at zero temperature.}
\label{fig:Damping}
\end{figure}

In contrast to the frequency, the damping rate of the radial
breathing mode reveals two transitions, at $\tilde{T}\simeq 0.5$
and at $\tilde{T}\simeq 1.0$, as shown in Fig.~\ref{fig:Damping}.

The lower  transition at $\tilde{T}=0.5$ is indicative of a
superfluid phase transition, consistent with our previous study of
damping versus temperature~\cite{Kinast}. Between $\tilde{T}=0.1$
and $\tilde{T}=0.5$, the damping rate decreases with decreasing
$\tilde{T}$, the opposite of the behavior expected for a
degenerate, collisionally hydrodynamic
gas~\cite{Vichi,Guery,BruunViscous} and consistent with a
superfluid picture~\cite{Kinast}. This behavior is similar to that
observed in the radial breathing mode of a
BEC~\cite{DalibardBreathing}. For $0.65\leq\tilde{T}\leq 1.0$, the
damping rate is nearly independent of temperature. This change in
behavior is accompanied by a significant, reproducible notch in
the damping rate near $\tilde{T}=0.6$. Using
Eq.~\ref{eq:NaturalTemp}, we find that $\tilde{T}=0.5$ corresponds
to $T/T_F=0.35$. This transition temperature is  somewhat higher
than the predicted superfluid transition temperature of
$T_c=0.29\,T_F$, as well as the value $T_c=0.27\,T_F$ estimated
from the observed slope change in the heat capacity, after
temperature calibration~\cite{JointScience}. However, it is not
clear that the observed change in the damping rate should occur
precisely at the same temperature as for the change in heat
capacity~\cite{JointScience}.

Damping below $\tilde{T}=0.5$ may arise from the  interaction
between a superfluid core and a normal gas in the edges of the
cloud. Normal fermionic excitations are present even when the core
is superfluid~\cite{LevinDensity}, because the local Fermi energy
at the edges can be smaller than $k_BT_c$. It is possible that the
 transition near $\tilde{T}=0.5$ arises as the core of the cloud changes from
superfluid to collisional  as the temperature increases. Indeed,
in the temperature region from $0.65\leq \tilde{T}\leq 1.0$, the
damping rate is close to the maximum allowed for a collisional gas
$1/\tau = 0.09\,\omega_\perp$~\cite{Guery}.

The higher temperature transition region, $1.0\leq \tilde{T}\leq
1.2$, may arise from the breaking of  noncondensed pairs. In this
region, the frequency remains near the unitary hydrodynamic value
while the damping rate rises roughly linearly with increasing
$\tilde{T}$ from approximately the maximum value allowed for a
collisional gas, $1/\tau = 0.09\,\omega_\perp$~\cite{Guery}, to a
value roughly 1.5 times larger, signaling the appearance of a new
channel of energy loss. The temperature range over which this
transition occurs is much smaller than the temperature scale, of
order $T_F$, over which the binary collision rate of a unitary gas
decreases in this temperature region~\cite{UnitaryCollisions}.
Although the frequency is nearly constant, this increase in
damping rate resembles that which we observed at low temperature,
at a magnetic field of 1080 G, where the coupling is reduced to
$k_F a = -1.35$~\cite{KinastMagDep}. Similar behavior was first
observed in Ref.~\cite{Bartenstein} and attributed to a pair
breaking process~\cite{Bartenstein,KinastMagDep}. The enhanced
damping in all of these experiments may be of the same nature:
both the decrease in coupling and the increase in temperature
reduce the pairing gap, making it comparable to the collective
excitation energy $\hbar\omega$ and therefore, causing fermion
pairs to break.

The second transition region occurs at high temperatures where the
energy of a strongly-interacting Fermi gas  merges with that of an
ideal noninteracting Fermi
gas~\cite{KinastScience,JointScience,ChenScience}. Using
Eq.~\ref{eq:NaturalTemp}, we see that the second transition region
$1.0\leq \tilde{T}\leq 1.2$, corresponds to $0.71\leq T/T_F\leq
0.86$, close to the  temperature range estimated for the vanishing
of noncondensed pairs~\cite{JointScience,ChenScience,ChenPrivate}.
In principle, merging of the ideal and unitary gas energies near
$T_F$ can arise (at least in part) because the unitary gas becomes
classical at high temperatures. However, the appearance of
enhanced damping supports the concept that the binding energy of
noncondensed pairs is decreasing  as part of the merging
process~\cite{JointScience,ChenScience}, causing collective
excitations to break pairs.

We have investigated the possibility that the observed variation
of the damping rate with temperature might arise in part from
oscillations of different components of the gas at different
frequencies. In a Bose-Einstein condensate with a thermal cloud,
such behavior leads to revivals of the net oscillation amplitude,
altering the apparent decay rates~\cite{AdamsDamping}. In the
present experiments, we find no evidence for such revivals, even
after increasing the time over which the decay of the mode is
observed.

The near linear dependence of the damping rate on $\tilde{T}$ in
the region $0.1\leq \tilde{T}\leq 0.5$ is well fit by
\begin{equation}
\frac{1}{\tau\,\omega_\perp}=0.146\,(0.004)\,\tilde{T}-0.0015\,(0.0014),
\label{eq:ExpLinearDamping}
\end{equation}
for the main data set taken at 4.6\% trap depth and $N=2\times
10^5$ atoms. The damping rate extrapolates close to zero at zero
temperature, consistent with a superfluid.

We have examined the dependence of the damping rate on the trap
oscillation frequency $\omega_\perp$ and on the number of atoms
$N$. Dimensional analysis requires that $1/\tau =
\omega_\perp\,f(T/T_F,N,\lambda)$, where $f$ is a dimensionless
function.

For fixed $T/T_F$ (or fixed $\tilde{T}$), we find that the
function $f$ cannot have a strong number dependence. For example,
it cannot be $\propto k_BT_F/(\hbar\omega_\perp)\propto N^{1/3}$
or its inverse. This is established by examining the scaling of
the damping rate with the atom number. We find that the damping
rate at  4.6\% of full trap depth does not change appreciably when
the number is reduced by a factor of $\simeq 3$: The damping rate
at reduced number, open circles in Fig.~\ref{fig:Damping}, lies
very close to the main data set (solid dots) when plotted versus
$\tilde{T}$. Hence, it is likely that $1/\tau$ depends on $N$ only
via the combination $T/T_F$, and the most general formula for
$1/\tau$ is then limited to
\begin{equation}
\frac{1}{\tau}=\omega_\perp\,f(\frac{T}{T_F},\lambda).
\label{eq:DampingGeneral}
\end{equation}
Experimentally, we are not able to test whether the damping rate
depends on $\lambda$. We have verified that $1/\tau$ versus
$\tilde{T}$ scales approximately as $\omega_\perp$ by monitoring
the breathing mode in the trap at 0.85\% of full depth
($\omega_\perp=728(4)$ Hz , squares in Fig.~\ref{fig:Damping}) and
at 19\% ($\omega_\perp=3343(20)$ Hz, diamonds). In both cases,
$1/\tau$ in units of $\omega_\perp$ is comparable to that of the
main data set.

Quantum viscosity~\cite{ShuryakQuantViscosity} recently has been
suggested as the mechanism for the small damping rate observed for
the axial breathing mode in Ref.~\cite{Bartenstein}. The quantum
viscosity $\eta$ is of order $\hbar k_F/\sigma$, where the
collision cross section $\sigma\propto 1/k_F^2$ in the unitary
limit~\cite{OHaraScience}. Hence, $\eta\propto \hbar\,
k_F^3\propto \hbar\, n$, where $n$ is the density.  For our
system,  the radial damping rate arising from quantum viscosity is
estimated to be $1/(\tau\omega_\perp)=3\times 10^{-5}$. This is
consistent with the extrapolated value at $T=0$, but cannot
explain the observed rates of order $0.014\,\omega_\perp$ at our
lowest temperature. Hence, the  decay of the radial mode is
probably by a different mechanism.


This research is supported by the Physics Divisions of the Army
Research Office  and the National Science Foundation, the Chemical
Sciences, Geosciences and Biosciences Division of the Office of
Basic Energy Sciences, Office of Science, U. S. Department of
Energy, and the Fundamental Physics in Microgravity Research
program of the National Aeronautics and Space Administration.


\begin{thebibliography}{42}
\bibitem[{Str()}]{Stringarishift}
\bibinfo{note}{This frequency shift was first derived by S. Stringari, private
  communication.}
\bibitem[{z2x()}]{z2x2}
\bibinfo{note}{We ignore here an additional small correction $\lambda^2\langle
  z^2\rho^2\rangle/(4\langle\rho^2\rangle)$. This changes the coefficient in
  Eq.~\ref{eq:FreqCorrection} from 2/5 to $2/5 \,+\, 1/16$. In the zero
  temperature Thomas-Fermi approximation, Eq.~\ref{eq:CorrectionTF}, the
  coefficient changes from $32/25$ to $32/25\, +\, 1/5$. The corresponding
  coefficient in the noninteracting gas correction~\cite{Kinast,KinastMagDep}
  is also increased from $6/5$ to $7/5$.}
\end{thebibliography}

\end{document}